# Actinides measurements on environmental samples of the Garigliano Nuclear Power Plant (Italy) during the decommissioning phase

A. Petraglia[1], C. Sirignano[1], R. Buompane[1], A. D'Onofrio[1], C. Sabbarese[1], A. M. Esposito[2], F. Terrasi[1]

[1]CIRCE, Dipartimento di Matematica e Fisica, Università degli Studi della Campania "L. Vanvitelli" Caserta, Italy
[2]SoGIN, Garigliano NPP, Sessa Aurunca (Caserta)

*An environmental survey was carried out in order to provide an adequate and updated assessment of the radiological impact that the decommissioning operations of the Garigliano NNP may have procured to the environment of the surrounding area. Some isotopes of uranium ($^{235}U$, $^{236}U$, $^{238}U$) and plutonium ($^{239}Pu$, $^{240}Pu$) and some γ-emitter radionuclides ($^{60}Co$, $^{137}Cs$ and $^{40}K$,) were measured to quantify the possible contamination and identify the origin source. Actinides isotopes were measured with the AMS technique that is able to detect elements in traces and reach sensitivity that cannot be obtained with other methods. The results show that the anthropogenic component is essentially due to the atmospheric fallout and no contamination can be charged to the NPP. Data are represented in geo-referenced maps to highlight the distribution area and some particular aspects of each measured radionuclide.*

## Introduction

The study of the radiological impact of a Nuclear Power Plant (NPP) decommissioning activities is important for the population and the environment of the area surrounding the plant, for the workers operating inside the plant and for characterisation and classification of the structural materials to be removed. Hence, a radiological characterization of environmental and structural materials is necessary. The Garigliano Nuclear Power Plant (GNPP), located in central Italy, about halfway between Rome and Naples, is in the decommissioning phase from 90s. The Centre for Isotopic Research on Cultural and Environmental Heritage (CIRCE, Caserta, Italy), in collaboration with SoGIN (Nuclear Plant Management Company), is carrying on a research program to provide a set of characteristic isotopic markers to assess the possible radiological impact that decommissioning operations at the GNNP (Figure 1) could have provided to the surrounding area. Previous surveys (Sabbarese et al., 2005; Petraglia et al., 2012) have been carried to assess the contamination levels over the years in order to safeguard the health of people and the environment and, also, to lower the level of risk perception among the population by means objective and verifiable scientific data.
In the above campaigns we did, natural and artificial gamma emitter radionuclides were measured. Since the amount of the Uranium isotopes ($^{235}U$, $^{236}U$, $^{238}U$) and Plutonium isotopes ($^{239}Pu$, $^{240}Pu$) and their abundance ratios can provide a signature of the contamination origin, in the actual campaign also these radionuclides were measured to obtain a very sensitive survey tool. These measurements require a method with high sensitivity. The Accelerator Mass Spectrometry (AMS) technique is the most sensitive available for the detection of long-lived radioisotopes at ultra-trace levels. At the CIRCE laboratory the AMS technique was extended to some long-lived radioisotopes and was applied for measuring the concentration and isotopic ratios of U and Pu isotopes in environmental and structural samples (De Cesare et al., 2011).
The analysis of the radionuclide quantity and their ratios could give much information. The $^{236}U/^{238}U$ ratio provides information on neutrons exposure. In fact, the $^{236}U$ is generated by neutron activation of $^{235}U$. Its presence in nature is minimal and its ratio with respect to $^{238}U$ is between $10^{-8}$ and $10^{-11}$. Higher values give us information about the presence of anthropogenic contamination (Steier et al., 2008).

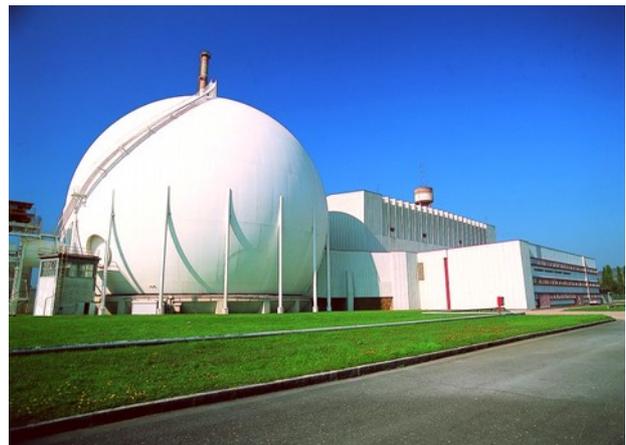

Figure 1. The Garigliano Nuclear Power Plant, Sessa Aurunca CE – Italy.

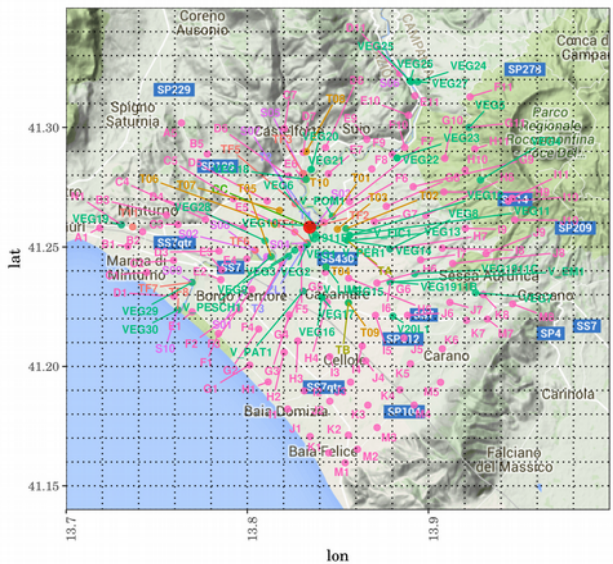

Figure 2. Location of the 191 environmental samples collected in the surroundings of the GNPP (in red).

Another, the analysis of $^{60}$Co, originating from neutron activation of $^{59}$Co, gives us information on the degree of contamination of the matrix from the reactor material, or it could also be an indicator of exposure to the neutron flow. The fission product $^{137}$Cs is an indicator of the degree of the sample contamination with radioactive anthropogenic material.

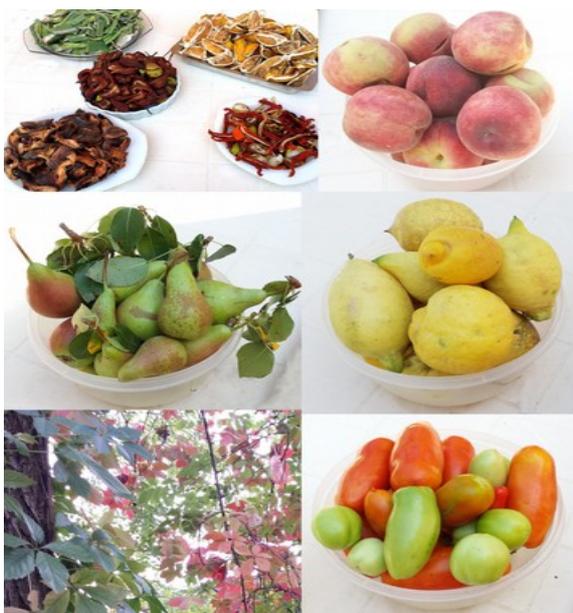

Figure 3. Montage of some of the 49 collected vegetal samples.

With the advancement of dismantling activities a new and comprehensive measurement campaign has become necessary in the years 2015-2017. It has been involved both structural materials and the environment around the plant. The first purpose has been to define the radiological signature (footprint) of the buildings and structures of the GNPP, so to guide future decommissioning operations and to monitor the different phases of the activities. The other purpose is an update and an in-depth study of the environmental matrix measurements in the area around the plant. Here, only the results of environmental samples are shown and discussed, while the results of structural materials are reported in a future work (Terrasi et al. 2017).

### Sampling area and method

The entire plain of Garigliano hosting the GNPP, from the sea at the foot of the surrounding hills, was chosen as survey area. Samples taken inside the GNPP area and from the surroundings were subject to different types of measurements for the quantitative determination of radionuclides with different characteristics, but all effective markers of the GNPP influence.

The environmental sampling points are shown on the map of the Figure 2. Soil samples (pink) were collected in points arranged according to a uniform square grid covering the Garigliano plain: the points were chosen approximately on each 1,5 km mesh. Some samples were also collected on the surrounding hills in easily accessible points. Seven other sampling points were selected in the Sele plain, approximately 130 km away from the GNPP, in order to have a background reference. Overall, no. 100 soil samples in the Garigliano plain and no. 7 samples in the Sele plain have been collected. The sampling was made by digging a 20x20x20 cm$^3$ hole in soil, whose surface is previously cleaned by any organic surface layer (usually herbaceous).

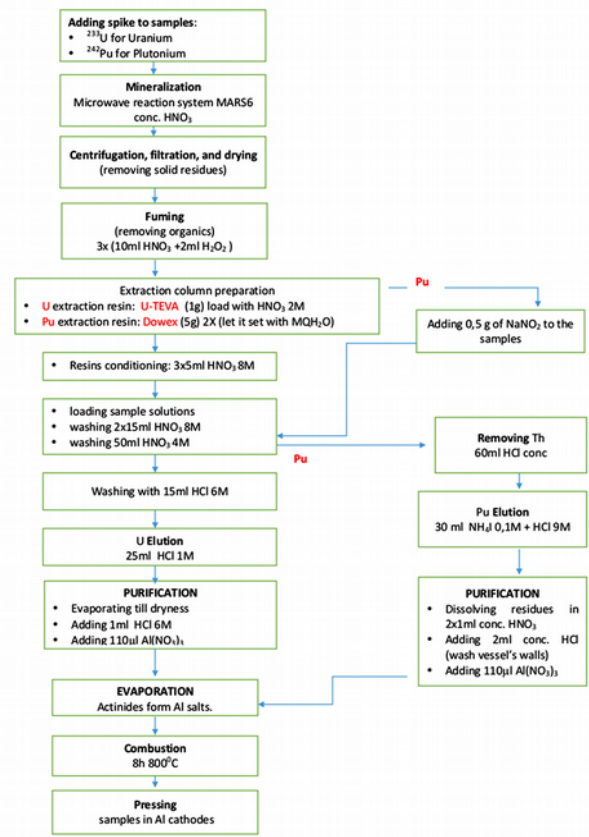

Figure 4: U and Pu chemical purification for AMS measurements.

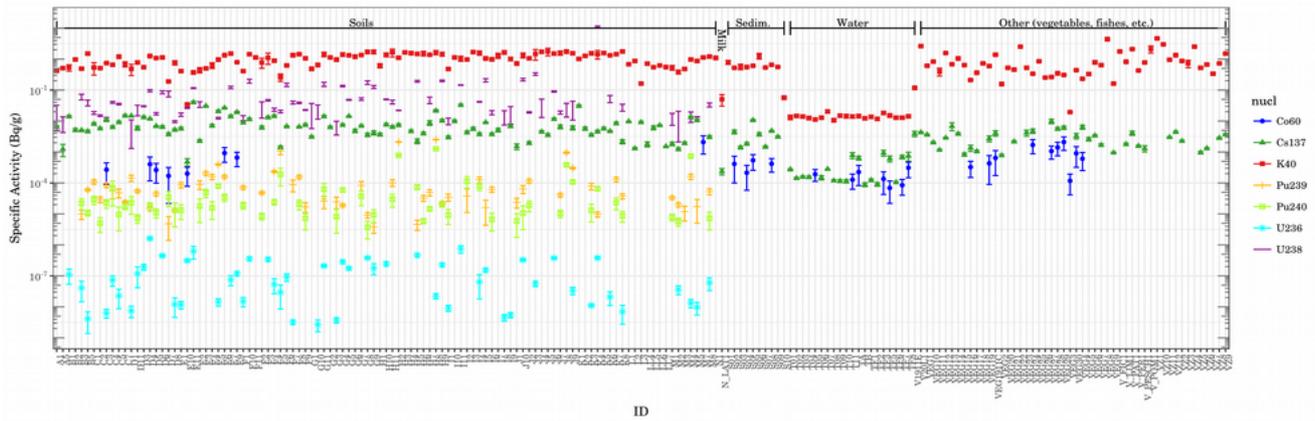

Figure 5. Synoptical comparison of the specific activity values of the three γ-emitter radionuclides and two actinides.

Plant samples (green labels in Figure 2) were taken from the native plants of the area, both wild and cultivated plants, leaves and fruits, depending on availability. In total the number of plant specimens is 49: tomatoes, ferns, ivies, figs, olive trees, peaches, lemons, pomegranates, potatoes, green beans and others. Along the river, no. 10 sediments and no. 7 water samples were taken; and another no. 11 groundwater samples were acquired from wells in the area immediately surrounding the GNPP. Finally, other different environmental samples were taken: fishes, mussels, milk, spring waters, etc. Altogether no. 191 environmental samples were collected; among them, no. 80 randomly selected soil samples were chosen to perform uranium and plutonium AMS measurements.

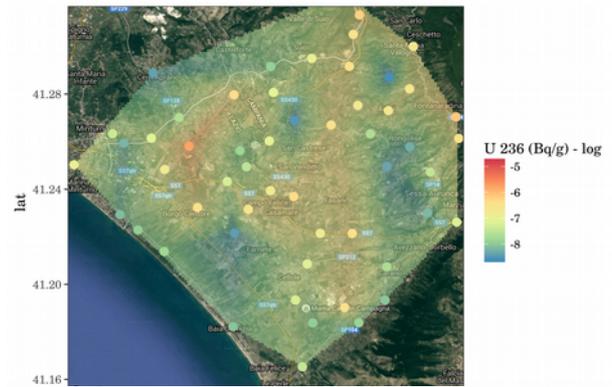

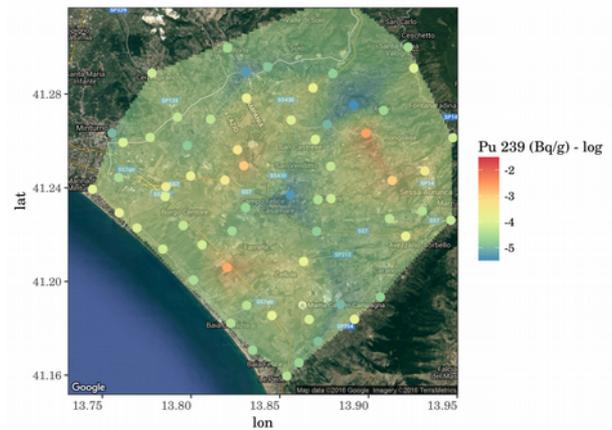

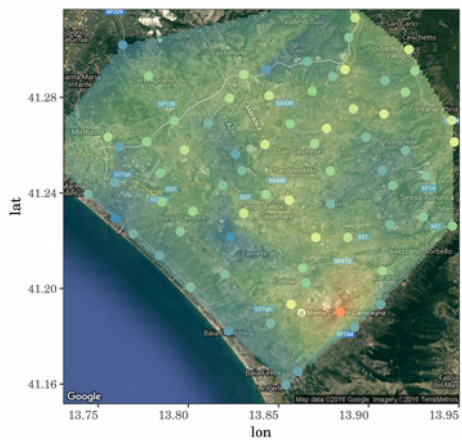

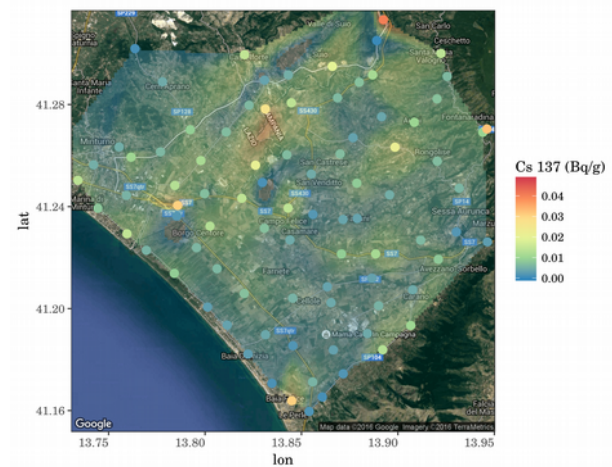

Figures 6a,b,c,d. False colour maps of specific activities of some radionuclides under study.

## Sample preparation and measurements

The γ measurements were carried out using a high-resolution germanium hyper-pure γ-ray detector (1.9 keV resolution at 1.332 MeV and 70% efficiency) properly shielded; spectra were acquired, displayed and analysed on computer running Ortec GammaVision. Before measuring, samples were dehydrated, homogenized, sieved and put in Marinelli vessels for the measurement. Each sample was measured with a long counting time (12-24 h) to allow the detection of small amounts of activity. From the analysis of the resulting spectrum, the $^{137}$Cs, $^{60}$Co (if present) were considered and the natural $^{40}$K was considered as a comparison term.

The extraction of U and Pu from sample was performed by means the chemical procedures schematically described in Figure 4. Before, the sample was dried in an oven at 110°C and homogenized.

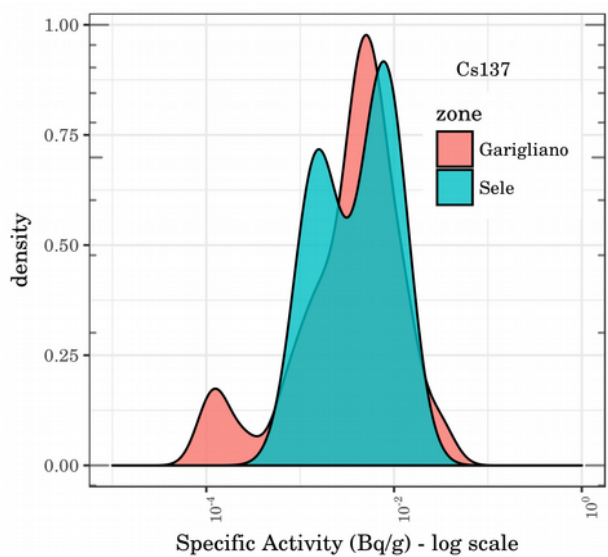

Figure 7. Distribution of the specific activities of $^{137}$Cs in soil samples collected in the Garigliano and Sele plains.

The ultra-sensitive actinides measurements were carried out by the AMS facility at the CIRCE, aiming to the determination of concentrations and isotopic abundances of U down to natural level and of Pu at ultra trace-level. The 3 MV CIRCE tandem accelerator is of Pelletron type. Since the installation of CIRCE in 2005, the measuring capabilities of the original system, which was then intended mainly for $^{14}$C dating, have been extended to allow U and Pu isotopes detection (De Cesare et al., 2013).

## Results

Figure 5 shows synoptically the comparison of the specific activity values (in logarithmic scale) of the γ-emitter radionuclides and actinides, grouped for each environmental matrix and coded with different colour and symbols: $^{60}$Co (blue), $^{137}$Cs (dark green), $^{40}$K (red), $^{236}$U (cyan), $^{238}$U (violet), $^{239}$Pu (orange) and $^{240}$Pu (light green).

It is possible to see that the specific activity values of $^{40}$K are, on average, one order of magnitude greater than those of $^{238}$U, which is one order of magnitude greater of the anthropogenic $^{137}$Cs. Other radionuclide have in general lower specific activities values.

The values of $^{236}$U ranges from $10^{-6}$ to $10^{-9}$ Bq/g.

It also emerges that the values of specific activities of soils, river sediments and plants are generally comparable among them, but higher than those of water samples, due to the low density of bulk material in water.

Figure 6 shows the maps of the distributions of some actinide isotopes ($^{236}$U, $^{238}$U and $^{239}$Pu) and of the $^{137}$Cs in the Garigliano plain (in logarithmic colour scale, except for Caesium). There is no area showing particularly high concentrations. In particular, no increase of any radionuclides concentration is seen in the area surrounding the NPP. Nevertheless, individual cases of extreme values can be seen, due to statistical distribution of these isotopes, which is expected to be log-normal.

The comparison of $^{137}$Cs specific activity values of soil samples collected in the Sele plain, 130 km southern from the plain under study, gives no significant differences, confirming the hypothesis of a global fallout origin of the anthropogenic contamination (Figure 7).

## Discussion and Conclusions

The samples collected around the GNPP have been prepared and measured for radiological signature with high sensitivity techniques. The results show values in agreement with the average of other environments not affected by nuclear power plants activities. The whole analyses show no significant alterations in the radiological characteristics of the area surroundings the plant, with an overall radioactivity depending mainly from the global fallout and natural sources; in particular, fallout of the Chernobyl accident plays a cardinal role (Alexakhin et al., 2007). Therefore, the values of anthropogenic radionuclides do not show any increase in concentrations in the plant area and have consistent values with the environmental contamination due to the fallout, in accordance with other studies (Quinto et al., 2009) and with the environmental campaigns previously carried out. This thesis is further confirmed by the mapping of the $^{236}$U and $^{238}$U specific activity. Indeed, the total specific quantity of $^{238}$U in the samples varies in a range of about two orders of magnitude (except for an outlier). This variability depends on the nature of soil. Similarly, the inspection of the $^{236}$U gives values coherent with global fallout.